\documentstyle[12pt,epsf]{article}
\renewcommand{\bar}[1]{\overline{#1}}

\begin{document}
 
\begin{flushright} 
SLAC-PUB-7314\\ 
BIHEP-TH-96-23\\
October 1996 
\end{flushright}
 
\bigskip\bigskip
\centerline{\Large \bf Asymmetric Quark/Antiquark Hadronization}
\smallskip\smallskip
\centerline{{\Large \bf in  $e^+e^-$ Annihilation} 
\footnote{\baselineskip=13pt
Work partially supported by the Department of Energy, contract
DE--AC03--76SF00515 and by National Natural Science Foundation of China
under Grant No.~19445004 and National Education Committee of China
under Grant No.~JWSL[1995]806.}}
\vspace{42pt}
 
\centerline{\bf Stanley J.~Brodsky}
\vspace{8pt}
\centerline{Stanford Linear Accelerator Center} 
\centerline{Stanford University, Stanford, California 94309, USA}
\centerline{e-mail: sjbth@slac.stanford.edu} 
\vspace*{8pt} 
\centerline{\bf Bo-Qiang Ma}
\vspace{8pt}
\centerline{CCAST (World Laboratory), P.O.~Box 8730, 
            Beijing 100080, China}
\centerline{Institute of High Energy Physics, Academia Sinica}
\centerline{P.~O.~Box 918(4), Beijing 100039, China} 
\centerline{and}
\centerline{Faculty of Natural Sciences, Kwanghua Academy of Sciences}
\centerline{Beijing 100081, China}
\centerline{e-mail: mabq@bepc3.ihep.ac.cn} 
\vfill 
\centerline{Submitted to Physics~Letters~B.} 
\vfill 
\newpage
 
$$ $$
$$ $$

\begin{center} 
{\Large \bf Abstract}
\end{center}
 
We point out that the fragmentation of a strange quark into nucleons
versus antinucleons is not necessarily identical $D_{p/s}(z,Q^2) \neq
D_{\bar p/s}(z,Q^2)$, even though the perturbative contributions from
gluon splitting and evolution are $ p \leftrightarrow \bar p$
symmetric. The observation of such asymmetries in the hadronization of
strange and other heavy quarks can provide insight into the
nonperturbative mechanisms underlying jet fragmentation in QCD.
 
\vfill
\centerline{PACS numbers: 13.65.+i, 11.30.Hv, 12.39.Ki, 13.87.Fh} 
\vfill\newpage
 
\section{Introduction}
 
A standard assumption often made in the analysis of sea quark
distributions is that the strange and antistrange contributions to the
nucleon structure functions are identical: $s(x,Q^2)= \bar s(x,Q^2)$.
Although the perturbative QCD contributions generated from gluon
splitting $g \to s \bar s$ and PQCD evolution do have this property,
the input distributions associated with the intrinsic bound-state
structure of the nucleon wavefunction need not be $s \leftrightarrow
\overline s$ symmetric. For example, the nonvalence quark
distributions associated with the $K \Lambda$ intermediate state, the
lowest mass meson-baryon fluctuation containing an intrinsic strange
quark-antiquark pair in the nucleon, have the property that the $s$
quark has a harder Bjorken $x$ distribution than the $\overline s$
antiquark\cite{Bro96}. Furthermore, due to the angular momentum
properties of the $ \left \vert K \Lambda \right\rangle $ intermediate
state, the $s$ quark helicity is antialigned with that of the parent
nucleon and the $\overline s$ helicity is unaligned. We have recently
presented a quantitative model of the strange and charm distributions
and their quark/antiquark asymmetries which is based on power-law or
Gaussian mass-weighted distributions of the interpolating meson-baryon
states in the nucleon light-cone Fock state wave function \cite{Bro96}.
The resulting asymmetry of the strange sea also allows the nucleon to
have additional nonzero $C=-1$ moments such as $s - \overline s$
contributions to the slope of the proton's charge radius and magnetic
moment.
 
Gribov and Lipatov \cite{GL1} have shown that the fragmentation
functions $D_{H/q}(z,Q^2)$ measured in quark and jet hadronization at $
z \simeq 1$ are related by crossing to the quark distributions
$G_{q/H}(x,Q^2)$ measured in deep inelastic scattering processes at $x
\simeq 1$. Thus any asymmetry between $G_{s/H}(x,Q^2)$ and $G_{{\bar
s}/H}(x,Q^2)$ implies a corresponding asymmetry in strange quark
fragmentation: $D_{H/s}(z,Q^2) \neq D_{H/{\bar s}}(z,Q^2)$ and $D_{
\bar H/s}(z,Q^2) \neq D_{H/s}(z,Q^2)$.
 
Thus it is possible that a strange quark will fragment differently into
a proton versus an antiproton. At first sight this seems antiintuitive
since if the strange quark produces a gluon which then hadronizes, the
resulting baryon and antibaryon products will be symmetric. If the $s$
quark hadronizes by coupling to a higher Fock component $\left \vert
uud s \bar s \right\rangle$ of the proton, then there will be no
asymmetry for $s \to p X$ versus $s \to \bar p X$ if the Fock state is
symmetric under $s \leftrightarrow \bar s$. The quark/antiquark
fragmentation asymmetry which we shall discuss in this paper is a
consequence of nonperturbative QCD processes where, for example, the
strange quark hadronizes via a $\Lambda$ in an intermediate state which
subsequently couples to a proton. Of course, the weak decay of the
strange quark is explicitly nucleon/antinucleon asymmetric so that
testing for such asymmetries requires that the weak decay products have
to carefully isolated from the hadronization processes. We discuss how
this can be done below.
 
\section{Anomalies of the Quark Sea}
 
The quark/antiquark asymmetry of the strange sea of the nucleon is
just one of a number of empirical anomalies relating to the composition
of the nucleons in terms of their nonvalence sea quarks: the European
Muon Collaboration has observed a large excess of charm quarks at large
momentum fraction $x$ in comparison with the charm distributions
predicted from photon-gluon fusion processes \cite{EMC82}; the large
violation of the Ellis-Jaffe sum rule observed at CERN \cite{EMC,SMC}
and SLAC \cite{E142,E143} indicates that a significant fraction of the
proton's helicity is carried by the sea quarks; the violation of the
Gottfried sum rule measured by the New Muon Collaboration (NMC)
\cite{NMC91} indicates a strong violation of flavor symmetry in the
$\bar u $ and $\bar d$ distributions; furthermore, there are
difficulties in understanding the discrepancy between two different
determinations of the strange quark content in the nucleon sea
\cite{CTEQ93,CCFR} assuming conventional considerations \cite{Ma96} and
perturbative QCD effects \cite{Glu96}.
 
The ``intrinsic" quark-antiquark ($q \bar q$) pairs generated by the
nonperturbative meson-baryon fluctuations in the nucleon sea
\cite{Bro96,Bro81,Sig87,Bur92,Pi} are multi-connected to the valence
quarks of the nucleon, and thus they have distinct properties from
those of the ordinary ``extrinsic" sea quarks and antiquarks generated
by QCD evolution. The concept of ``intrinsic charm" was originally
introduced \cite{Bro81} in order to understand the large cross-sections
for charmed particle production at high $x$ in hadron collisions, and a
recent comparison \cite{Vog96} between the next-to-leading order
extrinsic charm and the EMC charm quark structure function measurements
confirms the need for intrinsic charm. More recently, intrinsic $q \bar
q$ pairs of light-flavor $u$, $d$ and $s$ were studied using a
light-cone meson-baryon fluctuation model \cite{Bro96}. The model
predicts significant asymmetries in the momentum and helicity
distributions of the quarks and antiquarks of the nucleon sea and
provides a consistent framework for understanding the Gottfried and
Ellis-Jaffe sum rule violations and the conflict between different
measures of strange quark distributions.
 
Although the quark/antiquark asymmetries for the intrinsic $q \bar q$
pairs can explain a number of experimental anomalies \cite{Bro96},
there is still no direct experimental confirmation for such
asymmetries. The quark/antiquark asymmetries for the light-flavor $u
\bar u$ and $d \bar d$ pairs are difficult to measure since one has the
freedom to re-define the sea quarks by the requirements of sea
quark/antiquark symmetries due to the excess of net $u$ and $d$ quarks
in the nucleon. The quark/antiquark asymmetries for the heavy $b \bar
b$ and $t \bar t$ pairs are small due to the large $b$ and $t$ quark
masses. Thus the intrinsic strange $s \bar s$ and charm $c \bar c$
quark/antiquark asymmetries are the most significant features of the
model and the easiest to observe. It will be shown in this work that
the hadronic jet fragmentation of the $s$ and $c$ quarks in
electron-positron ($e^+e^-$) annihilation may provide a feasible
laboratory for identifying quark/antiquark asymmetries in the nucleon
sea.
 
\section{Asymmetric Quark/Antiquark Hadronization}
 
If there are significant quark/antiquark asymmetries in the
distribution of the $s \bar s$ pairs in the nucleon sea, corresponding
asymmetries should appear in the jet fragmentation of $s$ versus $\bar
s$ quarks into nucleons. For example, if one can identify a pure sample
of tagged $s$ jets, then one could look at the difference of
$D_{p/s}(z)-D_{\bar p/s}(z)$ at large $z$, where $D_{h/q}(z)$ is the
fragmentation function representing the probability for the
fragmentation of the quark $q$ into hadron $h$ and $z$ is the fraction
of the quark momentum carried by the fragmented hadron.
 
In the following we will discuss in some detail a test of the
strange/antistrange asymmetry of the nucleon sea using the strange
quark jets produced in $e^+e^-$ annihilation and then extend our
discussion to the case of charm quark jets. A key ingredient for such
tests is a source with unequal numbers of quark $q$ and antiquark
$\bar q$ jets in a specific direction so that we could identify a pure
$q$ (or $\bar q$) jet. Secondly, we require the identification of the
detected nucleon which is from the fragmentation of this quark $q$ (or
$\bar q$) jet rather than from other sources or subsequent processes.
At the $Z^0$-boson resonance in $e^+e^-$ annihilation, there is a
forward-backward asymmetry $A^f_{BF}$ in the production of fermion
pairs $f \bar f$. The asymmetries $A^f_{BF}$ for charged-leptons and
quarks (light-flavors, $c$, and $b$) have been measured to be of the
order 10\%, in agreement with the predictions of the Standard Model
\cite{PDG}. In case of polarized electron beam, the forward-backward
asymmetries are significantly enhanced. From this fact we know that
more $s$ (or $c$) quarks than $\bar s$ (or $\bar c$) quarks are
produced along the incident electron direction with left-handed
polarization. One can insure high $s \bar s$ purity by requiring a
$\phi$ meson to be directly produced at large momentum fraction $z$ in
the backward-going jet. This is a feature of fragmentation models and
is supported by experiment \cite{jets}: the $\phi$ at large $z$ arises
dominantly from the combination of the backward $\bar s$ quark with an
$s$ quark from a produced $s \bar s$ pair in the color-field. The
forward jet is then predominantly an $s$ quark. Alternatively, if the
$\phi$ is produced at large momentum fraction $z$ in the forward-going
jet, then the backward jet is predominantly an $\bar s$ quark.
 
In the meson-baryon fluctuation model of intrinsic $q \bar q$ pairs
\cite{Bro96}, the $s$ quark has a higher momentum fraction than the
$\bar s$ in the proton wavefunction. Thus we expect that protons are
produced at higher momenta from the $s$ quark hadronic fragmentation
compared to $\bar s \to p$ fragmentation. By using the high-$z$ $\phi$
tag, we can insure that the background from other $Z^0$ decays is
small. From symmetry considerations, we should have an excess of $s \to
p$ over $s \to \bar p$ at high momentum fraction in the forward going
jet with a backward high-$z$ $\phi$, and {\it vice versa}. Furthermore,
we can cut out protons fragmented at small $z$ to amplify the
fragmentation asymmetry. Thus the combination and self-consistency of
three aspects:

\begin{enumerate}
\item 
the asymmetry between the forward $D_{p/s}(z)$ and $D_{\bar p/s}$ with
a backward high-$z$ $\phi$;
\item 
the asymmetry between the backward $D_{\bar p/\bar s}(z)$ and $D_{p/
\bar s}$ with a forward high-$z$ $\phi$;
\item 
the asymmetry between the forward $D_{p/s}(z)$ with a backward high-$z$
$\phi$ and the backward $D_{p/\bar s}(z)$ with forward high-$z$ $\phi$
\end{enumerate} 
can provide a clean and unambiguous test of the strange-antistrange
asymmetry.
 
Any intermediate QCD hadronization process is allowed in the test of
hadronization asymmetries. However, we must avoid counting hadrons
which originate from weak decays, such as $s \to \Lambda X \to p X $
and $s \to \Sigma \to p X $, etc. Fortunately, the hyperons decay
through weak interactions with relatively very long life time, and thus
their decay products appear at long distances from the production
point. For example, the relativistic hyperon decays typically more than
$8~{\rm cm}$ away from the production point; thus it should be possible
to exclude protons from this source.
 
We now make a rough estimate of the magnitudes for the probability of
the fragmentation process $s \to p$ and the $s \to p$ versus $s \to
\bar p$ asymmetry for the forward going $s$ jet using the
Gribov-Lipatov reciprocity relation, which connects the annihilation
and deep inelastic scattering (DIS) processes in their physical regions
\cite{GL1,GL2,GL3}. The fragmentation function $D_{h/q}(z)$ for a quark
$q$ splitting into a hadron $h$ is related to the distribution function
$q_h(x)$ of finding the quark $q$ inside the hadron $h$ by the
reciprocity relation 
\begin{equation} 
D_{h/q}(z)=x \,q_h(x), \label{GL}
\end{equation} 
where $z=2 p \cdot q/Q^2$ is the momentum fraction of the produced
hadron from the $q$ jet in the annihilation process, and $x=Q^2/2 p
\cdot q$ is the Bjorken scaling variable in the DIS process. Although
there are corrections to this relation from experimental observation
and theoretical considerations\cite{GL3}, it can serve as a first
estimate of the fragmentation function. The strange quark distribution
in the proton, $s_p(x)$, has been evaluated in the light-cone
meson-baryon fluctuation model \cite{Bro96}, and the agreement with the
available data is reasonable. We thus use the calculated $s_p(x)$ in
\cite{Bro96} to evaluate the probability of the fragmentation $s \to
p$. We also make a minimum $z$ cut for the protons to reduce possible
contamination from the background. For $z_{\rm cut}=0.5$ we have
$\int_{z_{\rm cut}}^1 {{\rm d} z} D_{p/s}(z) \sim 10^{-4}$, and for
$z_{\rm cut}=0.6$ the value of $\int_{z_{\rm cut}}^1 {{\rm d} z} D_{p/s}(z)$
will be reduced by two additional orders of magnitude. Despite the
crudeness of the model and the limited range of the Gribov-Lipatov
reciprocity relation, we expect that the above estimate should serve as
a reasonable guide to the size of the expected effects.
 
\begin{figure}[htb]
\begin{center}
\leavevmode {\epsfxsize=3.5in \epsfbox{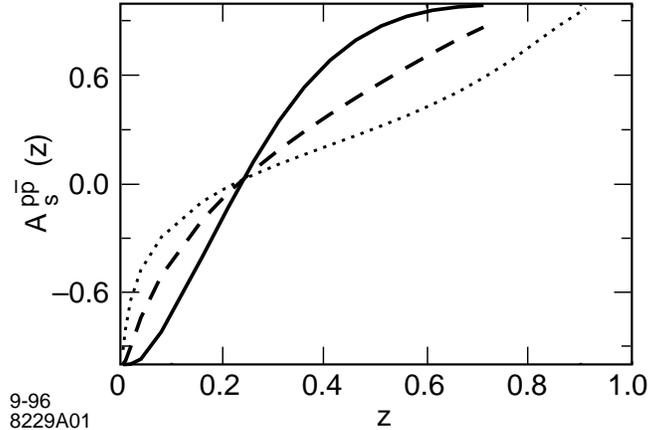}} 
\end{center}
\caption{\baselineskip 13pt 
The $s \to p$ versus $s \to \bar p$ asymmetry $A_s^{p\bar p}(z)$ as a
function of the momentum fraction $z$ of the produced proton
(antiproton) from the $s$ jet. The curves are the calculated results
for $A_s^{p\bar p}(z)$ in the light-cone meson-baryon fluctuation model
with Gaussian wavefunction $\psi_{{\rm Gaussian}}({\cal M}^2)=A_{{\rm
Gaussian}}\ {\rm exp} (-{\cal M}^2/2\alpha^2) $ in the invariant mass $
{\cal M}^2=\sum_{i=1}^{2} \ \frac{{\bf k}^2_{\perp i}+m_i^2}{x_i} $.
The full (dashed) curve is the calculated result of $A_s^{p\bar p}(z)$
$\alpha=330$ ($530$) MeV. The dotted curve is the result with a larger
$\alpha=800$ MeV adjusted for consistency with the available empirical
constraints on the strange-antistrange asymmetry in the proton
structure function.}
\label{epemf1}
\end{figure}

We define the $s \to p$ versus $s \to \bar p$ asymmetry
\begin{equation} 
A_s^{p\bar p}(z)=\frac{D_{p/s}(z)-D_{\bar p /s}(z)}{D_{p/s}(z)+D_{\bar
p /s}(z)}, 
\end{equation} 
which can be measured through the quantity
\begin{equation} 
A_s^{p\bar p}(z)=\frac{N[\phi~p(z)]-N[\phi~ \bar
p(z)]} {N[\phi~ p(z)]+N[\phi~ \bar p(z)]} / A^s_{BF},
\label{Ae}
\end{equation}
where $N [h_1~ h_2]$ represents the number of tagged $h_1$ and $h_2$
events under the specified kinematics in $e^+e^-$ annihilation. We
normalize the asymmetry to the value of the strange forward-backward
asymmetry $A^s_{BF}$ which can be measured by the tagged $\phi$ and
$\Lambda$ versus $\bar \Lambda$ (or strange versus antistrange mesons)
events or by tagging high momentum strange versus antistrange hadrons
with contributions from other flavors removed \cite{SFB}. The
combination of the $\phi$ tag on one jet with a strange  or antistrange
hadron tag on the other side jet can be used to identify the $s \bar s$
quark jets as well as identify which jet is $s$ or $\bar s$. The
asymmetry $ A_s^{p\bar p}(z)$ can be enhanced by restricting the
analysis to two-jet events since fragmentation from gluon jets will be
symmetric. Using the reciprocity relation Eq.~(\ref{GL}) we have
\begin{equation} 
A_s^{p\bar p}(z)=\frac{s_p(z)-\bar s_p(z)}{s_p(z)+{\bar s}_p(z)},
\end{equation}
from which we can evaluate the asymmetry $A_s^{p\bar p}(z)$ from the
strange-antistrange asymmetry $s(x)/\bar s(x)$ in Ref.~\cite{Bro96}. In
Fig.~\ref{epemf1} we present the calculated $A_s^{p\bar p}(z)$ in the
light-cone meson-baryon fluctuation model \cite{Bro96} with parameters
as those in Fig.~4 of Ref.~\cite{Bro96}. The strange and antistrange
quark distributions $s_p(x)$ and $\bar s_p(x)$ are evaluated by a
light-cone two-level convolution formula for the $\left \vert K \Lambda
\right\rangle$ component of the nucleon multiplied by a factor
$d_v(x)\vert_{\rm fit}/d_v(x)\vert_{\rm model}$ in the same framework to
reflect the effect of QCD evolution. We see that the calculated
$A_s^{p\bar p}(z)$ asymmetry has a strong $z$-dependence as a
consequence of the corresponding predictions of our model for a
significant quark-antiquark asymmetry in the nucleon sea. Thus given
sufficient statistics, there should be no difficulty in testing the
quark-antiquark asymmetry in the nucleon sea by measuring the predicted
$z$-dependence in the $s \to p$ versus $s \to \bar p$ asymmetry
$A_s^{p\bar p}(z)$ through Eq.~(\ref{Ae}).
 
\vspace{.5cm}
 
We can extend the above discussion to tests of the charm-anticharm
asymmetry in the nucleon sea; for example, one can define the $c \to p
+ X + c$ versus $c \to \bar p + X + c$ asymmetry in $e^+e^-$
annihilation at the $Z^0$ in analogy to $A_s^{p\bar p}(z)$. The
advantage of charm over strangeness for studying such asymmetries is
that:
\begin{enumerate}
\item 
The leading charm quark decays and any protons produced from open charm
weak decay can in principle be traced back to the source.
\item 
The protons associated with charmed meson charmed baryon fluctuations
will be fragmented at higher $z$ than the protons which come from
simple gluon fluctuations.
\end{enumerate} 
Such measurements are feasible for those $c$ (and $b$) decays where the
vertices from the $c$ and $\bar c$ decays are displaced and can be
measured. We predict a small but nonzero asymmetry for the
fragmentation distributions $D_{p/c}(z)$ versus $D_{\bar p/c}(z)$ from
the distribution functions $c_p(x)$ and $\bar c_p(x)$ in our model of
intrinsic $q \bar q$ pairs using the Gribov-Lipatov relation, in
analogy to the above discussion for the strange jets.
 
The probability of intrinsic charm quarks in the nucleon sea is
estimated \cite{Vog96} to be of the order of $0.6\pm 0.3 \% $, an order
of magnitude smaller than that of the strange quark $\sim 5\% $
\cite{Bro96}. Therefore we expect that the fragmentation function
$D_{p/c}(z)$ will be reduced by one order of magnitude compared to that
of $D_{p/s}(z)$. One can define the $c \to p + X + c$ versus $c \to
\bar p + X + c$ asymmetry in analogy to $A_s^{p\bar p}(z)$. It is
natural to expect a somewhat smaller $z$-dependence of the charm
asymmetry compared to that of $A_s^{p\bar p}(z)$.
 
\section{Conclusion}
 
Electron-positron annihilation, particularly at the $Z^0$-resonance,
can provide a laboratory for testing nucleon/antinucleon asymmetries
in the QCD hadronization of strange and charm quark jets. Such
asymmetries are characteristic of intrinsic $q \bar q$ pairs in the
nucleon wavefunctions. In particular, we predict a significant
$z$-dependence of the $s \to p$ versus $s \to \bar p$ asymmetry from
the hadronization of forward $s$ jets measured through tagged $\phi$
events under specified conditions. We also suggest a test of the
corresponding charm/anticharm hadronization asymmetry through the $c
\to p + X + c$ versus $c \to \bar p + X + c$ processes. The
experimental observation of hadronization asymmetries can provide
insight into nonperturbative mechanisms in jet fragmentation.

\section*{Acknowledgment} 
We would like to thank P.~Burrows, Jin Li, L.~N.~Lipatov, D.~Muller,
H.~Quinn, L.~Vitale, and Ji-Min Wu for helpful conversations.


\newpage

\bigskip 
\begin{thebibliography}{99}
 
\bibitem{Bro96} 
S.~J.~Brodsky and B.-Q.~Ma, Phys.~Lett.~{\bf B 381} (1996) 317.
 
\bibitem{GL1} 
V.~N.~Gribov and L.~N.~Lipatov, Phys.~Lett.~{\bf 37 B} (1971) 78;
Sov.~J.~Nucl.~Phys. {\bf 15} (1972) 675.
 
 
\bibitem{EMC82} 
EMC Collab., J.~J.~Aubert et al, Phys.~Lett.~{\bf B 110} (1982) 73.
 
\bibitem{EMC} 
EMC Collab., J.~Ashman {\it et al.}, Phys.~Lett.~{\bf B 206} (1988)
364; Nucl.~Phys.~{\bf B 328} (1989) 1.
 
\bibitem{SMC} 
SMC Collab., B.~Adeva {\it et al.}, Phys.~Lett.~{\bf B 302} (1993) 533;
{\bf B 357} (1995) 248; D.~Adams {\it et al.}, {\it ibid.} {\bf B 329}
(1994) 399; {\bf B 339} (1994) 332(E).
 
\bibitem{E142} 
E142 Collab., P.~L.~Anthony {\it et al.}, Phys.~Rev.~Lett.~{\bf
71} (1993) 959.
 
\bibitem{E143} 
E143 Collab., P.~L.~Anthony {\it et al.}, Phys.~Rev.~Lett.~{\bf 74}
(1995) 346; K.~Abe {\it et al.}, {\it ibid.} {\bf 75} (1995) 25.
 
\bibitem{NMC91} 
NM Collab., P.~Amaudruz {\it et al.}, Phys.~Rev.~Lett. {\bf 66} (1991)
2712; M.~Arneodo {\it et al.}, Phys.~Rev.~{\bf D 50} (1994) R1.
 
\bibitem{CTEQ93} 
CTEQ Collab., J.~Botts {\it et al.}, Phys.~Lett.~{\bf B 304} (1993)
159; H.~L.~Lai {\it et al.}, Phys.~Rev.~{\bf D 51} (1995) 4763.
 
\bibitem{CCFR} 
CCFR Collab., S.~A.~Rabinowitz {\it et al.}, Phys.~Rev.~Lett. {\bf 70}
(1993) 134; A.~O.~Bazarko {\it et al.}, Z.~Phys.~{\bf C 65} (1995) 189.
 
\bibitem{Ma96} 
B.-Q.~Ma, Chin. Phys. Lett. {\bf 13} (1996) 648 (hep-ph/9604342).
 
\bibitem{Glu96} 
M.~Gl\"uck, S.~Kretzer, and E.~Reya, Phys.~Lett.~{\bf B 380 } (1996)
171. See, however, V.~Barone {\it et al.}, Z.~Phys.~{\bf C 70} (1996)
83, for a contrary conclusion.
 
\bibitem{Bro81} 
S.~J.~Brodsky, P.~Hoyer, C.~Peterson, and N.~Sakai, Phys. Lett. {\bf B
93} (1980) 451; S.~J.~Brodsky, C.~Peterson, and N.~Sakai, Phys. Rev.
{\bf D 23} (1981) 2745.
 
\bibitem{Sig87} 
A.~I.~Signal and A.~W.~Thomas, Phys.~Lett.~{\bf B 191} (1987) 205.
 
\bibitem{Bur92} 
M.~Burkardt and B.~J.~Warr, Phys.~Rev.~{\bf D 45} (1992) 958.
 
\bibitem{Pi} 
E.~M.~Henley and G.~A.~Miller, Phys.~Lett.~{\bf B 251} (1990) 453.
 
\bibitem{Vog96} 
B. W. Harris, J. Smith, and R. Vogt, Nucl. Phys. {\bf B 461} (1996) 181.
 
\bibitem{PDG} 
Particle Data Group, L.~Montanet {\it et al.}, Phys.~Rev.{\bf D 50}
(1994) 1173, see, page 1353, note on the $Z$ boson and references
therein.
 
\bibitem{jets} 
See, e.g., W.~Hofmann, {\it Jets of Hadrons} (Springer-Verlag, Berlin
Heideberg, 1981).
 
\bibitem{GL2} 
G.~Schierholz, Phys.~Lett.~{\bf 47 B} (1973) 374.
 
\bibitem{GL3} 
DASP Collab., R.~Brandelik {\it et al.}, Nucl.~Phys.~{\bf B 148} (1979)
189; T.~Kawabe, Prog.~Theor.~Phys. {\bf 65} (1981) 1973.
 
\bibitem{SFB} 
DELPHI Collab., P.~Abreu {\it et al.}, Z.~Phys.~{\bf C 67} (1995) 1.
 
 
\nonfrenchspacing
\end{thebibliography}
\end{document}